\documentclass[prl,twocolumn,floatfix,showpacs]{revtex4}
\usepackage{epsfig}
\usepackage{bm}
\begin{document}

\title{Glassy Dynamics in the Adaptive Immune Response Prevents 
Autoimmune Disease}
\author{Jun Sun, David J. Earl, and Michael W. Deem}
\affiliation{Department of Bioengineering and
             Department of Physics \& Astronomy, Rice University,
Houston, TX 77005--1892, USA}

\begin{abstract}

The immune system normally protects the human host against death 
by infection. However, when an immune response is mistakenly 
directed at self antigens, autoimmune disease can occur. 
We describe a model of protein evolution to simulate the 
dynamics of the adaptive immune response to antigens. 
Computer simulations of the dynamics of antibody evolution
show that different 
evolutionary mechanisms, namely gene segment swapping and point mutation, lead 
to different evolved antibody binding affinities. 
Although a combination
of gene segment swapping and point mutation can yield a greater
affinity to a specific antigen than point mutation alone, the
antibodies so evolved are highly cross-reactive and would cause 
autoimmune disease, and this is not the chosen dynamics of the immune system. 
We suggest that in the immune system a balance has evolved 
between binding affinity and specificity in the mechanism
for searching the amino acid sequence space of antibodies.
\end{abstract}

\pacs{87.10.+e, 87.15.Aa, 87.17.-d, 87.23.Kg}

\maketitle


One of the main functions of the immune system
is to generate B cells that secrete antibodies, protein molecules 
that recognize and bind to antigen. The immune system has evolved a
hierarchical strategy to produce antibodies that
combat invading pathogens \cite{Janeway}.
The first level is VDJ recombination,
in which gene fragments are rearranged to encode functional
antibodies, 
producing a range of antibodies with a diversity on the order of $10^{12}$ to 
$10^{14}$. Upon invasion of the body by a foreign
antigen, the second level of B cell somatic hypermutation
evolution is initiated,
in which division, mutation, and selection 
of B cells occurs. Those B cells that produce antibodies that
bind the antigen with higher affinities are selected and propagated. 
Thus, the individual point mutations of somatic
hypermutation essentially perform an optimizing local search of amino
acid sequence space.

When an immune response is mistakenly 
directed at self antigens, autoimmune 
disease occurs~\cite{Janeway}.  
In most cases, the antibody/antigen interaction 
is highly specific. Sometimes, however, the antibody evolved in response
to one antigen can bind other antigens, and this phenomenon is 
called cross-reactivity~\cite{Goldsby}. Cross-reactivity happens 
when the other antigen has chemical features in common
with the original antigen and is quantified
experimentally by measuring the affinity of the antibody
for the other antigen.  Cross-reactivity is one mechanism
by which autoimmune disease may develop.

In this Letter, we present two major results. 
First, we show how different evolutionary mechanisms
influence the
relaxation dynamics of an evolving population of proteins. 
Specifically we study the dynamics of antibody generation 
by the immune system in response to 
pathogen invasion.  These dynamics are crucial to the
efficacy of the immune response.  
The hierarchical structure of our model 
protein Hamiltonian plays a critical role in the 
dynamics. The hierarchical structure also distinguishes our model 
from traditional short- or
long-range spin glass energy models~\cite{Fischer, Gross}. 
Second, we show that antibodies with significantly higher
affinities to antigen than those produced in a primary immune response
can be found with a biologically-plausible evolution process,
but are more cross-reactive and would greatly increase
susceptibility to autoimmune disorders.
The concept of cross-reactivity is related
to that of the chaos exponent in traditional spin glasses~\cite{Young, Ritort},
except that for cross-reactivity we are  interested 
in the immediate energy response rather than in the
equilibrated response of the system to a change in the
couplings.
Taking our two results together, we show that 
in the primary adaptive immune response a careful
balance has evolved between
specificity for and binding affinity to foreign antigen.


To express the interaction energy between antibodies and antigen, we 
use the generalized $NK$ model, a model that captures
important features of the immune response to vaccination and disease 
and of protein molecular evolution~\cite{Bogarad,Deem,Earl}.
The energy of interaction for a given protein or antibody sequence
is defined in the model as
\begin{equation}
U=\sum_{i=1}^M U_{\alpha_i}^{\rm sd} + \sum_{i>j=1}^{M} U_{ij}^{\rm sd-sd}
         +\sum_{i=1}^{P} U_i^{\rm c} \,
\label{Hamiltonian}
\end{equation}
The term $U^{\rm sd}$ includes 
interactions within a secondary structure.
Interactions between secondary structures are
essential to protein folding and function. These
terms are in $U^{\rm sd-sd}$.
The total number of interactions with a typical amino acid
is roughly 12, and half of these are in $U^{\rm sd}$ ($2 (K-1)$), and
the other half in $U^{\rm sd-sd}$ ($D (M-1)/N$).
The term $U^{\rm c}$ is the direct 
binding interaction between the antibody and antigen. 
The number of antibody secondary structural subdomains is $M=10$,
and the number of antibody amino acids involved directly in the binding
is known to be approximately $P=5$ \cite{Bogarad}. We 
define the binding constant between antibody and antigen
to be equal to $K=\exp^{a-b \langle U \rangle}$ where $a$ and $b$ are constants 
found by comparison to immunological data~\cite{Deem}. Thus, the lower 
the overall energy, the better the binding affinity of the antibody 
to the given antigen. 
The secondary structure energy $U^{\rm sd}$ is 
\begin{equation}
 U_{\alpha_i}^{\rm sd}=\frac{1}{\sqrt{M(N-K)}} \sum_{j=1}^{N-K+1}
      \sigma_{\alpha_i}(a_j,a_{j+1},\cdot \cdot \cdot,a_{j+K-1}) \,
\label{U1}
\end{equation}
where $N=10$ is the number of amino acids in a subdomain, and $K=4$ 
\cite{Kauffman}
specifies the range of the local interactions within the secondary
structure. We consider
five chemically distinct amino acid classes (negative, positive, polar,
hydrophobic, and other), and all subdomains 
belong to one of $L=5$ different types (helices, strands, loops, turns, 
and others). The quenched Gaussian random number $\sigma_{\alpha_i}$ is 
different for each value of its argument for a given subdomain type, 
$\alpha_i$. All of the Gaussian $\sigma$ values have zero mean and unit 
variance. The energy of interaction between secondary structures $i$ and $j$ is
\begin{eqnarray}
U_{ij}^{\rm sd-sd}=\sqrt{\frac{2}{DM(M-1)}} 
\sum_{k=1}^{D}\sigma_{ij}^{k}(a_{l_1}^{(i)}, 
\ldots, a_{l_{K/2}}^{(i)}; \nonumber \\
a_{l_{K/2+1}}^{(j)}, \ldots, a_{l_{K}}^{(j)})
\label{U2}
\end{eqnarray}
where $D=6$ specifies the number of interactions between 
secondary structures. 
Here  $\sigma^k_{ij}$ and the interacting amino acids, $l_1,...,l_K$, are 
selected at random for each interaction  $i, j, k$. The chemical binding 
energy of each antibody amino acid to the antigen is given by 
$U^{\rm c}_i= \sigma_i(a_i)/\sqrt{P}$. The contributing amino acid, $a_i$, and 
the unit-normal weight of the binding, $\sigma_i$, are chosen at random. 

In our first study, we replicate the immune system dynamics.
Each protein sequence is of the length of $ N \times M=100$ amino acids,
corresponding to the length of the heavy chain variable region 
of an antibody. Initially there are
$10^3$ sequences, as in the human immune response~\cite{Deem}. 
Two different strategies are used in our simulations to 
search protein sequence space for high affinity antibodies. 
The first strategy mimics the normal adaptive humoral immune
response. It starts with the combinatorial joining of optimized 
subdomains, after which the sequences undergo rounds of point mutation (PM) and 
selection. This is a simulation of the 
VDJ recombination, somatic hypermutation, 
and clonal selection that occurs
in B cell development~\cite{Goldsby}. We generate five 
optimized subdomain pools, each composed of $300$ low-energy subdomains, 
corresponding to the $L=5$ types. 
We use three sequences from each subdomain pool
in our VDJ recombination, mimicking the known diversity~\cite{Deem}.

In our second study, we include a more powerful search of sequence
space in the antibody evolution dynamics.
VDJ recombination is used to generate an initial
population, as in the normal
immune response, but in addition to somatic hypermutation, we perform
gene segment swapping (GSS) during
each round of evolution and selection.
GSS-type processes are used by experimentalists to produce antibodies
with binding constants $\approx 10^{11} - 10^{13}$ l/mol \cite{Schier}, and
exist within the natural hierarchy of evolutionary events \cite{Kidwell,Earl}.

In the process of GSS, a subdomain belonging to one of the 
five types is replaced by another one from the optimized subdomain pool of 
the same type. Each sequence undergoes an average of $0.5$ point mutations
in each round of selection in both strategies. In GSS+PM,
each subdomain in a sequence has a 
probability of $0.05$ of being replaced in the process of GSS.
Following mutation, selection occurs, and the 20\% lowest energy
antibodies are kept and are amplified to form the population
of $10^3$ sequences for the next round of mutation and selection.
The primary response is comprised of $30$ rounds of affinity maturation, 
corresponding to a lag phase of approximately 10 days~\cite{Deem}, 
during which B cells undergo clonal selection in response to antigen 
and differentiate into plasma cells and memory cells. 
The results that follow are averaged over $5000$ instances
of the ensemble. 

\begin{figure}
\begin{center}
\epsfig{file=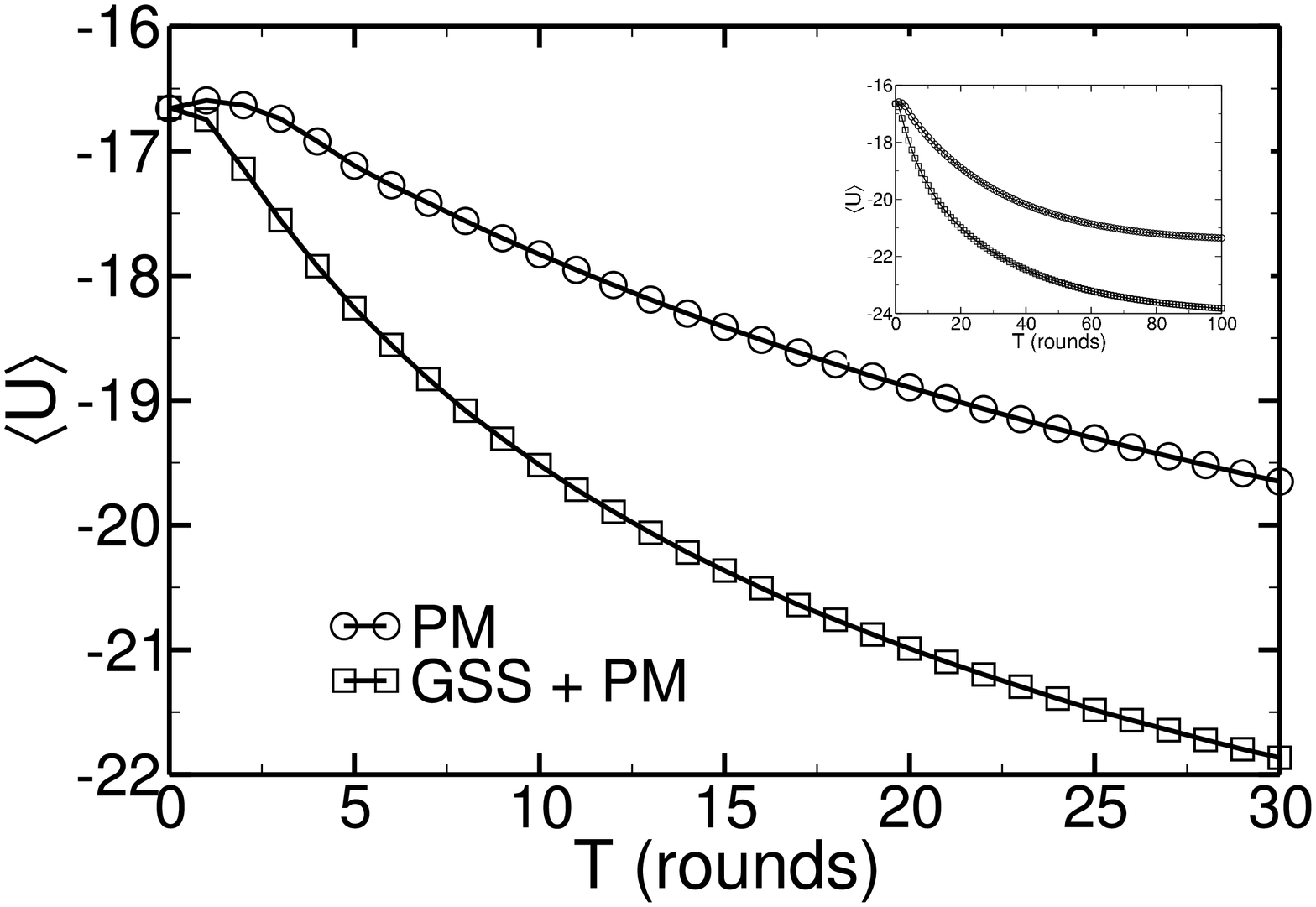,width=0.90\columnwidth,clip=}
\end{center}
\caption{ 
Evolution of the affinity energy for the cases of point mutation (PM) only 
and gene segment swapping (GSS) plus PM as a function of the number of rounds
of mutation and selection used to evolve the population of antibodies.
Energies evolved at larger number of rounds shown in inset.}
\label{fig:UandT}
\end{figure}

\begin{figure}[tbp]
\begin{center}
\epsfig{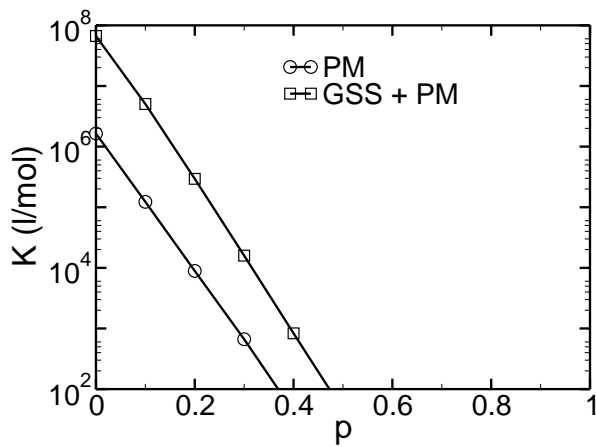}
\end{center}
\caption[]{Affinity of memory antibody sequences after a primary 
immune response for the two different immune system strategies 
(PM and GSS+PM) to altered antigens. 
The binding constant is $K$, and the antigenic distance 
of the new altered antigen from the original antigen is $p$. 
Cross-reactivity ceases at larger distances in the 
GSS+PM case (no cross-reactivity for $p>0.472$) than in the PM only 
case (no cross-reactivity for $p>0.368$).}
\label{fig:KandP}
\end{figure}

The average affinity of the population of antibodies evolves 
during mutation and selection. The evolution 
of the affinity energy under the two different strategies is 
shown in Fig.~\ref{fig:UandT}. 
The GSS+PM yields sequences 
with a lower energy than those from PM.
The gap is due
to each of the subdomains having an energy improved
 from an average to an exceptional one,
$\Delta U \approx 10 (U^{\rm sd}_{\rm best} - U^{\rm sd}_{\rm avg}) = -0.8$,
and due to better sd-sd interactions found by GSS.
In other words, the former is a better mechanism than the 
latter in searching sequence space
for higher affinity antibodies to the given antigen.  The 
best binding energy averaged over 5000 instances of the
ensemble is $U^{G}=-21.9$ 
in the GSS+PM case and $U^{P}=-19.7$ in the PM only case at 
round $30$, corresponding to affinities of $K=6.7 \times 10^7$ l/mol and 
$K=1.6\times 10^6$ l/mol, respectively. So, GSS+PM yields an 
affinity more than an order of magnitude improved, which is even better than 
the affinity obtained from a secondary response by PM only~\cite{Deem}. 
That is, the multi-spin flips of GSS+PM especially
accelerate the optimization of $U^{\rm sd-sd}$ in amino acid sequence space.
Antibodies 
with a higher affinity to the antigen work more effectively in many ways.
For example they can neutralize bacterial toxins, inhibit the infectiousness 
of viruses, and block the adhesion of bacteria to host cells at lower
serum concentrations of antibody~\cite{Goldsby}. 
Based solely on Fig.\ \ref{fig:KandP}, it is hard to understand why Darwinian
evolution did not result in GSS+PM or any other more
efficient strategy, rather than somatic hypermutation,
as the preferred strategy for B cell expansion.

A calculation of cross-reactivity, which quantifies
the specificity of the antibody, is used to compare the antibodies
generated by the two strategies. 
Thirty rounds of primary response
affinity maturation are conducted for both PM and GSS+PM.
The antigen is then changed by 
$p$, and so each interaction parameter in the Hamiltonian 
is changed with probability $p$. The affinity constants 
are calculated for the new antigen in the two cases. As shown 
in Fig.~\ref{fig:KandP}, we find that cross-reactivity ceases approximately 
$\Delta p=0.10$ later in the GSS+PM case. 
We also find that the affinity constant, $K$, decreases exponentially
with the degree of antigenic change, \emph{i.e.}\ the
binding energy, $U$, increases linearly
with the change in the antigen, $p$. 
Within the region of specific 
binding ($K>100$ l/mol), affinity is always better in the GSS+PM case.
These cross-reactivity 
experiments show that the antibodies generated by the dynamics of GSS+PM 
can recognize more antigens and with higher affinity than those
given by the PM dynamics only.  Such cross reactivity has
recently been observed \cite{Holler}.
Interestingly, $U(p=1) =  U(T=0) / 5$, since at $p=1$ the only correlation
is that $\alpha_i$ remains unchanged 20\% of the time.
The fraction $p$ of new
$\sigma$ values in $U^{\rm sd-sd}$ and $U^{\rm c}$
are centered around zero and negligible compared to the 
original, negative evolved values.
In $U^{\rm sd}$, a fraction $p$ of the $\alpha_i$ are changed,
and the original, evolved sequence gives a value of roughly
zero in a changed type $\alpha_j$.
The overall energy lost in $U$ is, thus, proportional to $p$,
Fig.\ \ref{fig:KandP}

\begin{figure}[tbp]
\begin{center}
\epsfig{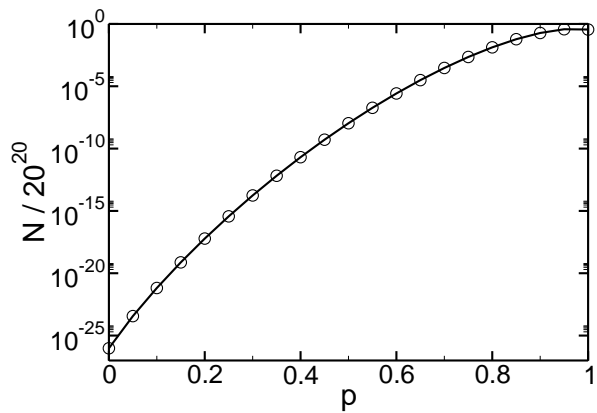}
\end{center}
\caption[]{The number of possible epitopes that are
an antigenic distance of $p$ from another epitope. 
The epitopes are assumed to be 20 amino acids in length. 
The plot is normalized, so that $\sum_p
N(p) = 20^{20}$.}
\label{fig:NandP}
\end{figure}

Now the question we must answer is how many protein molecules are
there between $p_1=0.368$ and $p_2=0.472$ that 
can be recognized by the antibodies produced through
 VDJ recombination and GSS+PM, 
but cannot be recognized by the antibodies produced 
through VDJ recombination and PM only. 
Antibodies recognize and bind to epitope regions of an antigen, so we
are interested in how many epitopes these two different classes 
of antibodies are likely to recognize.
We now show that the abnormal antibodies recognize $10^3$ 
times more epitopes than do the antibodies produced by the normal primary
immune response.
We take the typical epitope length to be
$B=20$ amino acids~\cite{Goldsby,Enrique}.
In the theory, the antigenic
distance is defined by the chemical composition  of the epitope.
The total number 
of possible epitopes is, therefore,
 $20^B$ since there are $20$ different amino 
acids~\cite{Campbell}. In Fig.~\ref{fig:NandP} we show 
the normalized number of epitopes that are at an
antigenic distance of $p$ from the native epitope of the antibody,
given by $N(p) = 19^i B!/[i! (B-i)!]$,  where $i =  p B$.
This number is $N(p_1) / 20^B =2\times 10^{-12}$  and
$N(p_2) / 20^B =2\times 10^{-9}$.
The number of epitopes between 
$p=0$ and $p_1$  and the number between $p_1$ and $p_2$  is approximately
$A_1 \approx 2 \times 10^{-12}$ and $A_2 \approx 2\times 10^{-9}$,
respectively, since $N$ grows exponentially with $p$.
In the limit of large $B$, $N(p_2) / N(p_1) \sim 
\exp \{ - B [ \ln (1-p_2) / (1-p_1) + p_1 \ln 19 (1/p_1-1) 
-      p_2 \ln 19 (1/p_2-1) ] \} $.  This ratio is $10^3$ for 20 amino 
acids.  The ratio varies between 160 and 4900 at the 15--25 amino acid
limits of known epitope sizes.  The ratio varies
by only 30\% with $\pm 10$\% variation of both $p$ values,
the precision to
which the generalized $NK$ model $p$ values agree with experiment
\cite{Deem,flu2}.

In order to answer the question of why selection has favored 
the relatively slower 
dynamics in the immune response, we consider the function of
the immune system as a whole rather than just the function of
a single antibody binding to a single
foreign antigen. The immune system is a delicate 
balance of a strong response to invading non-self antigens and 
weak or no response
to self antigens. Thus, it is very important for the 
antibodies produced by the immune system
 to discriminate self from non-self~\cite{Percus, Detours, Suzler}. 
An immune system incapable of recognizing an invading pathogen and 
initiating a response is an inadequate defense mechanism. On the other hand, 
production of antibodies binding to self antigens results in autoimmune 
diseases, \emph{e.g.}\ diseases such as
type I diabetes and rheumatoid arthritis.
It might be thought that GSS+PM would be superior dynamics, if only
the immune system were able to perform $\approx$ 10 rounds.
However, each subsequent exposure to related antigen would lead to another
$\approx$ 10 rounds.
After a couple such exposures, GSS+PM has evolved an
an unacceptably large binding constant, which saturates after several
exposures to of
the order of $10^{10}$ l/mol, whereas PM saturates at
the value of $10^7$ l/mol.
Even with PM dynamics, our model predicts that chronic infection
may lead to autoimmune disease, a mechanism
postulated to be responsible for some
fraction of rheumatic diseases, including arthritis
 \cite{Repo}.  To resolve the significance of this mechanism,
it would be interesting to see whether the distribution of onset times
is broad, as predicted by our model.

It has been experimentally shown that cross-reactivity can lead to 
autoimmune disease~\cite{Janeway}. 
From Figs.~\ref{fig:KandP}--\ref{fig:NandP}, we know that the antibodies 
obtained at the end of a primary immune response can recognize 
a random epitope with a length of $20$ amino acids with a probability 
of $\approx 10^{-12}$. It is also known that in a typical cell
there are $\approx 10^4$ proteins, each with a length of 
$\approx 500$ amino acids~\cite{Tan}. Antibodies binding
to proteins recognize only surface epitopes.  The amino acids 
exposed on the surface of proteins are typically part of a loop
or turn, and typically 1/3 of the loop or turn is exposed.
Thus, a typical contiguous recognition fragment length is 6--7 amino acids.
Given a typical length of 20 amino acids for the entire loop or turn,
and the typical epitope size of 20 amino acids, an
antibody will recognize roughly 3 non-contiguous regions of length
6--7 amino acids in the protein sequence of 500 amino acids.
There are approximately
$500^3 \approx 10^8 $ such epitopes.  
Given the $10^4$ proteins in a cell, there will be $\approx 10^{12}$ 
total epitopes expressed
in each cell. Thus, the number of epitopes recognized
by a typical antibody in each cell is $A_1 \times 10^{12} \approx 1$.
Even taking an exceptional protein of length 1000 amino acids, the
number increases only to 10.

This remarkable result shows that the antibodies produced by the
immune system recognize on average only their intended target.
Conversely, 
antibodies that would be evolved through an immune response 
composed of VDJ recombination 
followed by a period of GSS+PM would 
recognize on average $A_2 \times 10^{12} \approx 10^3$ epitopes in each cell.
Such antibodies, while having higher affinities for the intended
target, would also lead to 
many more instances of autoimmune disease.  Such promiscuous
antibodies would place too large a burden
on the regulatory mechanisms that eliminate the occasional
aberrant antibody \cite{Larche}.
Thus, we find that selection has successfully evolved the human 
immune system to generate antibodies that recognize on average
only the intended epitope after a humoral immune response.
Inclusion of ``more efficient'' moves is generally excluded 
by the bound $A \times 10^{12} = O(1)$.


\bibliography{autoimmune}

\end{document}